\documentclass{aa}  

\usepackage{graphicx}
\usepackage{txfonts}

\usepackage{booktabs}

\usepackage[colorinlistoftodos,color=blue!20]{todonotes}

\usepackage{ulem}
\usepackage{multirow}
\usepackage{longtable}

\usepackage{hyperref}
\hypersetup{colorlinks=true, linkcolor=blue, citecolor=blue, filecolor=blue, urlcolor=blue, pdftitle= Astrophysics”; http://www, pdfauthor=E389, pdfsubject=, pdfkeywords=}

\def\arcsec{\hbox{$^{\prime\prime\,}$}}
\def\deg{\hbox{$^\circ$}}
\def\Msun{\hbox{\,M$_{\odot}$\,}}

\def\kms{\,km~s$^{-1}$}
\def\Rsun{\hbox{\,R$_{\odot}$\,}}
\newcommand\Rtx{R_{\rm TX~Psc}}

\begin{document}

   \title{ALMA observations of the ``fresh'' carbon-rich AGB star TX\,Piscium}

   \subtitle{The discovery of an elliptical detached shell}

   \author{M.Brunner
          \inst{1}
          \and
          M.~Mecina
          \inst{1}
          \and
           M.~Maercker
          \inst{2}
          \and
          E.A.~Dorfi
          \inst{1}
          \and
          F.~Kerschbaum
          \inst{1}
          \and
          H.~Olofsson
          \inst{2}
          \and
          G.~Rau
          \inst{3,4}
                   }

   \institute{Department for Astrophysics, University of Vienna,
              T\"urkenschanzstrasse 17, A-1180 Vienna\\
              \email{magdalena.brunner@univie.ac.at}
         \and
          Department of Space, Earth and Environment, Chalmers University of Technology, 43992 Onsala, Sweden
          \and 
          NASA Goddard Space Flight Center, Code 667, Greenbelt, MD 20771, USA
          \and 
          Department of Physics, The Catholic University of America, Washington, DC 20064, USA
   }

   \date{Received Month nr, YYYY; accepted Month nr, YYYY}

 
  \abstract
   {}
   {The carbon-rich asymptotic giant branch (AGB) star TX~Piscium (TX~Psc) has been observed multiple times during multiple epochs and at different wavelengths and resolutions, showing a complex molecular CO line profile and a ring-like structure in thermal dust emission. We investigate the molecular counterpart in high resolution, aiming to resolve the ring-like structure and identify its origin.}
   {Atacama Large Millimeter/submillimeter Array (ALMA) observations have been carried out to map the circumstellar envelope (CSE) of TX~Psc in CO(2--1) emission and investigate the counterpart to the ring-like dust structure.}
   {We report the detection of a thin, irregular, and elliptical detached molecular shell around TX~Psc, which coincides with the dust emission. This is the first discovery of a non-spherically symmetric detached shell, raising questions about the shaping of detached shells.}
   {We investigate possible shaping mechanisms for elliptical detached shells and find that in the case of TX~Psc, stellar rotation of~2\,km/s can lead to a non-uniform mass-loss rate and velocity distribution from stellar pole to equator, recreating the elliptical CSE. We discuss the possible scenarios for increased stellar momentum, enabling the rotation rates needed to reproduce the ellipticity of our observations, and come to the conclusion that momentum transfer of an orbiting object with the mass of a brown dwarf would be sufficient.}


   \maketitle

\section{Introduction}

At the low- to intermediate-mass end of the final stages of stellar evolution, Sun-like stars evolve into asymptotic giant branch (AGB) stars, exhibiting a rapid increase in size and luminosity. This leads to the formation of a pulsating stellar atmosphere, where dust grains are formed and strong stellar winds originate \citep[e.g.][]{Habing1996}. A complex interplay between multiple physical processes is involved in this stage of stellar evolution, challenging our current attempts to reconstruct these processes in consistent models \citep{Hofner2018}. As a consequence of the significant amount of mass lost through stellar winds, huge circumstellar envelopes (CSEs) of molecular gas and dust are formed. Those CSEs can be analysed observationally to achieve insights into the mass-loss history of AGB stars, and therefore to disentangle some parts of the interplay of processes at work during the creation of these CSEs. In theory, the stellar wind is spherically symmetric and homogeneous, but due to variations in the mass-loss rate and possible irregularities imposed by intrinsic stellar properties, interaction with the surrounding (interstellar or previously ejected) medium, interaction with binaries, or possibly also the interaction with magnetic fields, this is in most cases not true. Therefore, the morphology and dynamics of CSEs rather represent snapshots of the recent mass-loss history of their host stars.

An example of the manifestation of the stellar mass-loss history in CSEs is the existence of so-called detached shells, currently found around about a dozen carbon AGB stars. These detached shells are believed to be linked to thermal pulses (TPs), which are short events (a few hundred years) of significantly increased mass loss, produced by alternating extinction and re-ignition of the hydrogen- and helium-burning shells in the stellar interior \citep[e.g.][]{Steffen2000,Mattsson2007TP,Olofsson1988,Olofsson1990}.

Observations of extended CSEs can be conducted in three different wavelength regimes: for the analysis of the dusty component, optical observations of scattered light on dust grains \citep[e.g.][]{Gonzalez-Delgado2001,Olofsson2010,Maercker2014} or infrared observations of the thermal dust emission \citep[e.g.][]{Cox2012,Jorissen2011} can be used. The second component, the molecular gas, can be observed in the millimeter and sub-millimeter wavelength range, and mostly the very stable and abundant CO molecule is used as a tracer of molecular CSEs \citep[e.g.][]{Olofsson1993}. Especially within the last decade, observatories such as the \textit{Herschel} Space Observatory, operating in the infrared, and the Atacama Large Millimeter/sub-millimeter Array (ALMA)\footnote{\url{https://almascience.nrao.edu}.}, operating in the millimeter/sub-millimeter range, have enabled high-resolution observations of previously unresolved or barely resolved CSEs and structures within them.

The target of this publication, TX~Piscium (\object{TX Psc}), is a relatively well-studied carbon-rich AGB star of spectral type C7,2 \citep{Yamashita1972}, which was originally classified as a variable star of type Lb in the General Catalogue of Variable Stars (GCVS) but \citet{Wasatonic1997} found an average period of 224\,days, changing the classification to a semi-regular variable (SRa/b). For a detailed overview of the derived and modelled stellar parameters, we refer to Table\,3 in \citet{Klotz2013}, and in the current work we adopt an effective temperature $T_{\mathrm{eff}}$ of TX~Psc of 3000\,K (average of values from multiple references in the literature), a luminosity $L$ of 7700\,L$_{\odot}$~\citep{Claussen1987}, and the most recent distance estimate to the star is $275^{+34}_{-26}$\,pc \citep{van-Leeuwen2007}. Regarding its evolutionary status, the average C/O ratio derived from multiple observations and models, as summarised by \citet{Klotz2013}, is 1.07, which indicates a recent transition from an oxygen-rich atmosphere to a carbon-rich one and makes TX~Psc a relatively "fresh" carbon star. Mass-loss rates reported in the literature range from $9.8 \times 10^{-7}$\,M$_{\odot}$/yr \citep[average value of][]{Olofsson1993} to $5.6 \times 10^{-7}$\,M$_{\odot}$/yr \citep{Loup1993}. We summarise the most important stellar parameters that we use in the course of this publication in Table\,\ref{tab:starparams}.

The CSE of TX~Psc has been observed at multiple wavelengths with different methods of observations, ranging from UV to sub-millimeter and radio wavelengths, including both low- and high-resolution observations. Asymmetries and clumps in the close vicinity of the star have been detected at small spatial scales in the optical and infrared \citep[e.g.][]{Cruzalebes1998,Ragland2006,Hron2015}, focusing on the dusty component of the CSE. The molecular counterpart has also been investigated with multiple instruments in the millimeter and sub-millimeter range \citep[e.g.][]{Heske1989,Olofsson1993,Schoier2000,Schoier2001}, but the spatial resolution of these single-dish observations has not been sufficient to map the CSE in detail. \citet{Heske1989} put limits on the spatial extent of the molecular CSE and report CO(1--0) and CO(2--1) emission at offsets of ~20\arcsec from the star. Because of the multi-component line profile of the CO lines, up to now no radiative transfer modelling has been attempted by any of the investigators, and suggestions about a bipolar or binary-shaped nature of the molecular emission have been raised in the literature. The first large-scale image of the CSE around TX~Psc with sufficient resolution to analyse the CSE structure was delivered by the \textit{Herschel}/PACS instrument \citep{Poglitsch2010} through the Mass loss of Evolved StarS (MESS) programme \citep{Groenewegen2011}. \citet{Jorissen2011} report the detection of a circular ring-like structure around the star with a radius of $\sim$17\arcsec as well as clear indication for a separate interstellar medium (ISM) interaction front.

In this paper, we present high-resolution ALMA Cycle 3 observations (PI: M. Brunner, project ID 2015.1.00059.S) of the CO(2--1) emission around TX~Psc, for the first time resolving the  molecular CSE in detail and revealing an elliptical detached shell at the location of the dusty ring-like structure. We discuss the implications of this discovery on the mass-loss evolution of the star as well as on the evolution and properties of AGB stars in general. 

In Sect.\,\ref{sec:observations} we describe the observations, followed by their subsequent analysis in Sect.\,\ref{sec:analysis}. Our results are discussed with respect to previous observations, stellar parameters, and evolutionary status of TX~Psc in Sect.\,\ref{sec:discussion}, and we summarise and conclude our findings in Sect.\,\ref{sec:conclusions}.

\begin{table}
\caption{Stellar parameters of TX~Psc.}
\begin{center}
\begin{tabular}{llrcc}
\toprule
\multicolumn{2}{c}{Stellar parameters}  \\
\multicolumn{2}{c}{TX~Psc}  \\
\midrule

Distance & $275^{+34}_{-26}$\,pc\,\tablefootmark{a} \\
Effective temperature & 3000\,K\,\tablefootmark{b}  \\
Gas mass-loss rate & $ 3.2\times 10^{-7}$ \,M$_{\odot}$/yr\,\tablefootmark{c}\\
Dust/gas mass ratio & $0.72\times 10^{-3}$\,\tablefootmark{c}  \\
Luminosity & 7700\,L$_{\odot}$\,\tablefootmark{d} \\
Stellar mass & 2\,M$_{\odot}$\,\tablefootmark{b} \\
Gas expansion velocity & $\sim$10\,km/s\,\tablefootmark{e} \\
Stellar velocity (LSRK) & $\sim$13\,km/s\,\tablefootmark{e}\\

\bottomrule
\end{tabular}
\end{center}
\tablefoot{Stellar parameters extracted from following references:\\
\tablefoottext{a}{\citet{van-Leeuwen2007}}
\tablefoottext{b}{\citet{Klotz2013}}
\tablefoottext{c}{\citet{Bergeat2005}}
\tablefoottext{d}{\citet{Claussen1987}}
\tablefoottext{e}{this study.}
}

\label{tab:starparams}
\end{table}

\section{Observations}
\label{sec:observations}

The ALMA observations were carried out in Cycle 3 (October 2015 -- September 2016) and consist of main array (MA), Atacama Compact Array (ACA), and total power array data, to recover all spatial scales larger than $\sim$1\arcsec. We requested an angular resolution of 1\arcsec to observe the CO(J = 2--1) spectral line in a mosaic of 45\arcsec $\times$ 45\arcsec around the star. This covers the ring structure seen in the IR dust emission, but not the close-by ISM interaction region. The spectral setup was optimised for the CO(2--1) line emission at 230.538\,GHz with a spectral resolution of 499.84\,kHz (0.65\kms). Two additional spectral windows were set up at 244.936 and 231.221\,GHz to cover the frequencies of the CS(5--4) and $^{13}$CS(5--4) lines, respectively, with the same spectral resolution as for the CO line. Additionally, one continuum spectral window was set up at 248.000\,GHz with a spectral resolution of 0.977\,MHz and a bandwidth of 1875\,MHz.

Standard calibration of the interferometric data was performed with the Common Astronomy Software Application (CASA)\footnote{\url{https://casa.nrao.edu}}, using Uranus and Neptune as flux calibrators, the quasars J2253+1608, J0006-0623, and J2232+1143 as bandpass calibrators, and J0006-0623 as phase calibrator. The total power data were calibrated according to standard procedure using the quasars J2327+0940, J0010+1058, and J2323-0317 as calibrators.
Imaging of the interferometric data was carried out and improved in collaboration with the Nordic ARC Node\footnote{\url{https://www.oso.nordic-alma.se}} using CLEAN. The requested angular resolution for this project was 1\arcsec, calculated for the requested signal-to-noise ratio (S/N) in the extended CSE. The MA baselines at the date of the observations allowed a higher angular resolution and we used uv-tapering to reduce the angular resolution and gain S/N in the faint regions of the CSE. Nevertheless, we also imaged the MA data with the full available spatial resolution (without uv-tapering) for the analysis of the brighter regions of the CSE.

Since the primary beams of the MA and ACA are different, due to the different antenna size, a previously unforeseen and undocumented problem prevented us from doing any useful array combination in the visibility plane: if the imaged area contains emission at the edges of the mosaic, which is the case for a multitude of large-scale structures to be observed with ALMA, and especially true for TX\,Psc with its close-by ISM interaction region, the different primary beams of the MA and ACA pick up different fractions of the signal from the edges of the mosaic. Therefore, they do not contain the same amount and spatial distribution of emission. Hence, not only is flux lost but also artefacts, which decrease the image quality of a combined image, are introduced. For this reason, we refrained from combining the MA and ACA data and in the following always analyse the data separately, focusing on the small-scale emission and fine structure of the observed CSE when using the MA data, and on the large-scale emission and general shape of the emission when using the ACA data. The slightly resolved total power data are analysed on their own as well, to confirm the general shape of the CSE.
The beam sizes and image properties for all datasets are given in Table\,\ref{tab:obsproperties}.

\begin{table}
\caption{Beam sizes and image properties of the ALMA observations.}
\begin{center}
\begin{tabular}{lcccc}
\toprule
Data 
&\multicolumn{1}{r}{\parbox{1.3cm}{\centering $\theta_\mathrm{b}$ \\{(arcsec)}}}
& \multicolumn{1}{l}{\parbox{1.0cm}{\centering PA\\{(\deg)}}}
& \multicolumn{1}{l}{\parbox{1.0cm}{\centering $\theta_{\mathrm{max}}$\\{(arcsec)}}}
& \multicolumn{1}{l}{\parbox{1.0cm}{\centering rms\\{(mJy)}}}\\

\midrule

MA & 1.06\,$\times$\,0.98 & -80.09 & 5.6 & 3\\
ACA & 7.60\,$\times$\,4.44 & -89.26 & 18.6 & 9 \\
Total power & 28.15 & -55.35 & -- & 44 \\


\bottomrule
\end{tabular}
\end{center}
\tablefoot{Given are the beam size $\theta_\mathrm{b}$, beam position angle PA, the maximal recoverable scale $\theta_{\mathrm{max}}$, and the average rms noise of line-free channels.}
\label{tab:obsproperties}
\end{table}

\section{Analysis}
\label{sec:analysis}

\subsection{Detached shell}

\begin{figure}[htbp]
   \centering
   \includegraphics[width=0.4\textwidth,trim={1.5cm 0cm 1cm 0cm}]{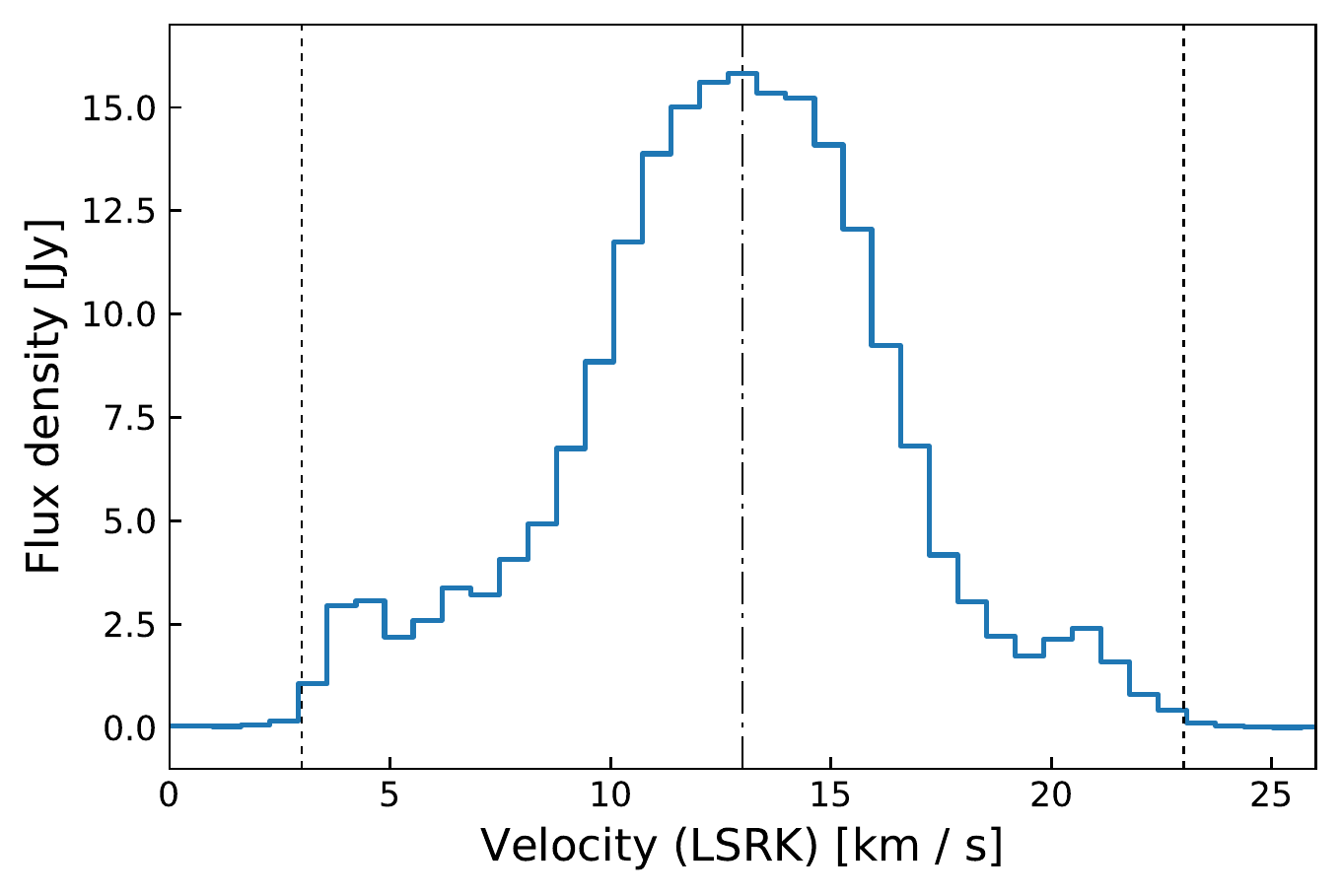} 
   \caption{CO(2--1) spectral line observed with the ALMA main array. The spectrum is extracted from a circular aperture of 25\arcsec radius. The black dotted-dashed line marks the stellar velocity of 13\,km/s (LSRK) and the dotted lines mark the stellar velocity $\pm$10\,km/s.}
   \label{fig:MAspec}
\end{figure}

\begin{figure*}[htbp]
   \centering
   \includegraphics[width=1.0\textwidth,trim={0cm 5.3cm 0cm 6.3cm}]{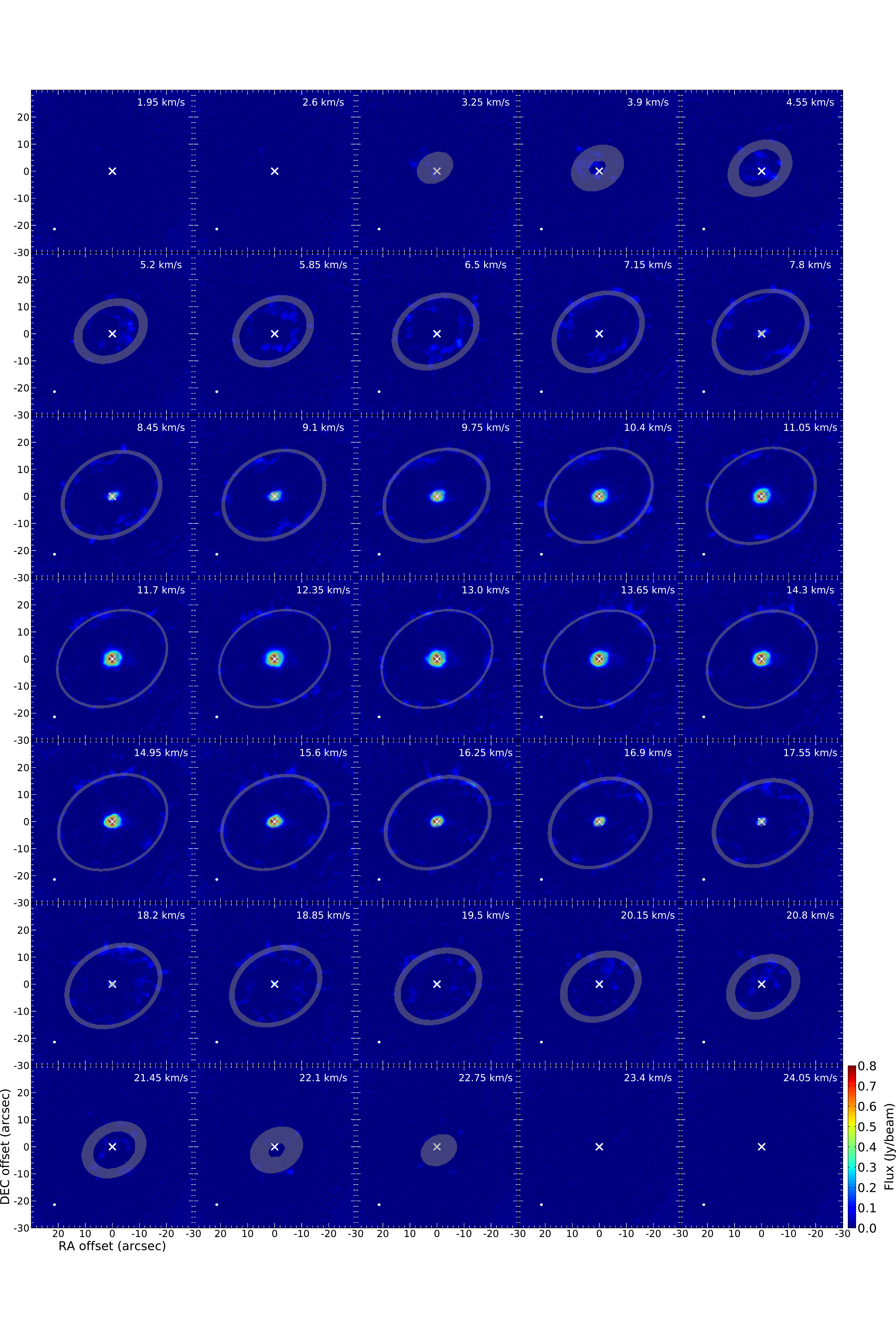} 
   \caption{Channel map of the CO(2--1) emission observed with the ALMA main array (colourscale) overlaid with the best fitting \textit{Shape} geometric model with an inclination of 100\deg and a PA of 30\deg (grey). North is up, east is left. The stellar position is marked as a white cross for reference. The beam is given as a white ellipse in the lower left of the panels.}
   \label{fig:MAShape}
\end{figure*}

\begin{figure*}[htbp]
   \centering
   \includegraphics[width=1.0\textwidth,trim={0cm 5.3cm 0cm 6.3cm}]{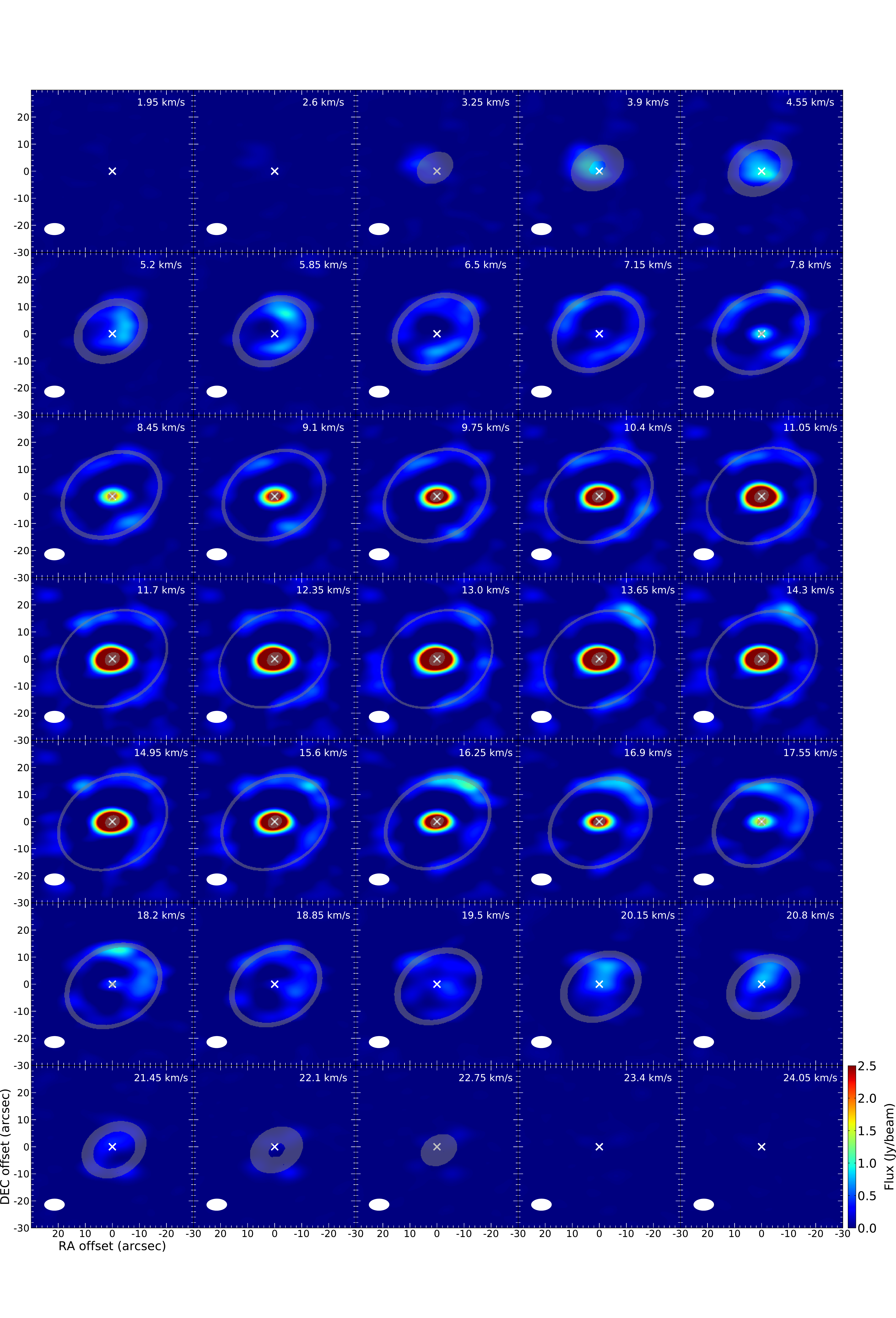} 
   \caption{Channel map of the CO(2--1) emission observed with the ACA (colourscale) overlaid with the best fitting \textit{Shape} geometric model with an inclination of 100\deg and a PA of 30\deg (grey). North is up, east is left. The stellar position is marked as a white cross for reference. The beam is given as a white ellipse in the lower left of the panels.}
   \label{fig:ACAShape}
\end{figure*}

We report the detection of a thin detached shell around TX~Psc, which is seen both in the high resolution MA and low resolution ACA data, as presented in the respective CO(2--1) channel maps in Figs.\,\ref{fig:MAShape} and \ref{fig:ACAShape}. The shell is inhomogeneous and clumpy, and not spherically symmetric but of elliptical shape in the plane of the sky. We measure an extent in the north-west/south-east direction of $\sim$22\arcsec (semi-major axis), while the extent in the north-east/south-west direction is only $\sim$17\arcsec (semi-minor axis), which results in an axis ratio of 1.3. The shell thickness seems to be roughly of the same size as the resolving beam, but since it appears to be very inhomogeneous, we cannot reliably constrain it. For the course of the subsequent analysis, we estimate a shell thickness of 1\arcsec. From the spectrum extracted from the image cube (Fig.\,\ref{fig:MAspec}) we determine an expansion velocity of roughly 10\,km/s for the shell. We assume that this velocity corresponds to the expansion along the semi-major axis, which is based on the assumption that the extent of the ellipsoidal shell along the line of sight is similar to the semi-major axis, in other words, that the ellipsoid is axisymmetric to the semi-minor axis and not significantly tilted with respect to the line of sight. With an expansion velocity of 10\,km/s along the semi-major axis of the shell, the upper limit of the age of the detached shell is $\sim$2600\,years.

To analyse the geometry using the projected ellipticity of the detached shell, we constructed a three-dimensional (3D geometrical model with \textit{Shape} \citep{Steffen2011} and compared rendered model channel maps with the observations to extract geometric parameters. The \textit{Shape} channel maps are created from a geometric model of homogeneous density with attached velocity field. As measured from the observations under the assumption of either no or very small inclination of the axisymmetric ellipsoid with respect to the line of sight, the model ellipsoidal shell has a semi-major axis of 22\arcsec, a semi-minor axis of 17\arcsec, and a thickness of 1\arcsec. With the assumption that the elliptical shell was generated during a thermal pulse (or any other single event happening at a specific time), the expansion velocity along the semi-minor axis has to be lower than along the semi-major axis, to generate an ellipsoidal shape. We describe the ellipsoidal velocity profile by an elliptical function, assigning 10\,km/s as the  velocity along the semi-major axis, the extracted maximal expansion velocity from the spectrum. The corresponding expansion velocity along the semi-minor (polar) axis then equals 7.7\,km/s. \textit{Shape} allows the rendering of channel maps, which we can directly compare to our observations.

Since the maximum projected ellipticity of an axisymmetric ellipsoid is encountered when the line of sight is perpendicular to the symmetry axis, we first rendered the 3D model in this configuration with an inclination angle of 90\deg and a position angle (PA) of 0\deg, and subsequently changed first the PA and later the inclination, to find a best fitting interval of both parameters to the observations. We arrived at a good fitting   PA interval of roughly 15 - 45\deg, with the best fit at a PA of approximately 30\deg, and a good fitting inclination interval of 90\deg - 110\deg, with the best fit at an inclination of approximately 100\deg. In this model the PA is counted anti-clockwise from north and an inclination angle $>90$\deg means that the bottom (or southern) half of the semi-minor axis is tilted towards the observer. Figure\,\ref{fig:MAShape} shows the MA channel maps overlaid with the rendered geometric model from \textit{Shape} with the best fitting PA and inclination.

The geometric projection effect of a rotational ellipsoid with corresponding velocity field viewed under an inclination of 100\deg~can best be seen in the extreme velocity channels, where the projected ellipse is located off-centre, shifted to north-east in the blue-shifted channels, and shifted to south-west in the red-shifted channels. Additionally, in those extreme channels the ellipse is thicker than in the channels around the systemic velocity (13\,km/s). The 3D model ellipsoid fits the observations very well in the velocity channels from around 7.8\,km/s to 18.2\,km/s and gradually deviates more from the observations when going to the more extreme velocities, where the observed shell looks more irregular and particularly clumpy. Figure\,\ref{fig:ACAShape} shows the ACA channel maps overlaid with the best \textit{Shape} model. For this lower resolution data, the agreement of the geometric model with the observations also holds for the more extreme velocities, where the MA data appears too patchy to constrain the geometry well.

\subsection{Present-day wind}

\begin{figure*}[htbp]
   \centering
   \includegraphics[width=1.0\textwidth,trim={0cm 0.5cm 0cm 1cm}]{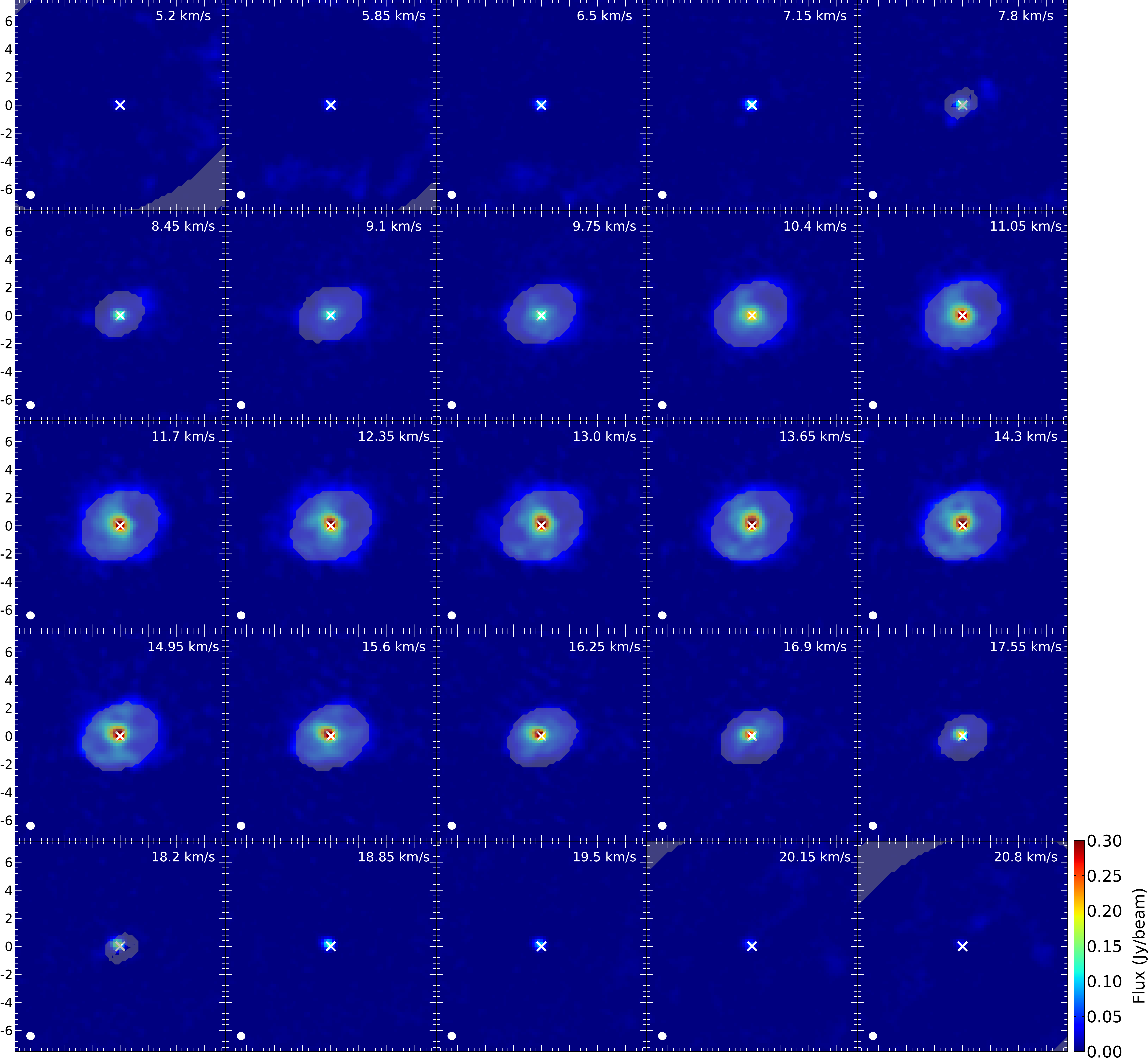} 
   \caption{ Channel map of the CO(2--1) present-day wind emission observed with the ALMA main array (colourscale), overlaid with the \textit{Shape} model (grey). North is up, east is left. The stellar position is marked with a white cross. The beam has a size of 0.61\arcsec $\times$ 0.57\arcsec and is given as a white ellipse in the lower left of the panels.}
   \label{fig:MApdmap}
\end{figure*}

Figure\,\ref{fig:MAShape} shows that the present-day wind is well resolved and the S/N is higher than in the shell. To analyse the present-day wind in detail, we used the images created without uv-tapering (as described in Sect.\,\ref{sec:observations}), using the full resolution of the main array with a beam of 0.61\arcsec $\times$ 0.57\arcsec and zoom in on the central 7.5\arcsec of the emission (Fig.\,\ref{fig:MApdmap}). The brightest and most compact part of the emission is roughly contained within the innermost 2\arcsec (in diameter). The extended part of the present-day emission also appears to be of elliptical shape with a size of roughly 3.25\arcsec along the semi-major axis (north-west/south-east direction) and roughly 2.5\arcsec along the semi-minor axis (north-east/south-west direction). Thus, the axis ratio of the present-day extended emission of 1.3 is the same as for the elliptical shell. It should be noted that the measurement uncertainty is quite high, since the exact extent of the present-day wind is hard to extract from the images due to inhomogeneities and its small size, compared to the image resolution. The extended present-day emission, as seen in the spectrum, roughly ranges from 7.2\,km/s to 18.9\,km/s, corresponding to an expansion velocity of about 5.9\,km/s along the line of sight. With these measurements, we model the geometric 3D structure and velocity profile of the present-day wind and confirm a good fit of the models to the observations (Fig.\,\ref{fig:MApdmap}). The expansion velocity along the semi-major axis is assumed to be the line of sight velocity, 5.9\,km/s, as a first approximation, and the expansion velocity along the semi-minor axis is 4.5\,km/s.  

Within the extended and elliptical present-day emission there seems to be a small cavity or at least an emission minimum to the north-west of the star, which is best visible at the velocity channel of 11.05\,km/s. In general, the extended present-day emission shows substructure at the resolution limit.

\subsection{Comparison of gas and dust}
\label{subsec:dust}
\begin{figure}
 \centering
   \includegraphics[width=0.5\textwidth,trim={0cm 2cm 0cm 2cm}]{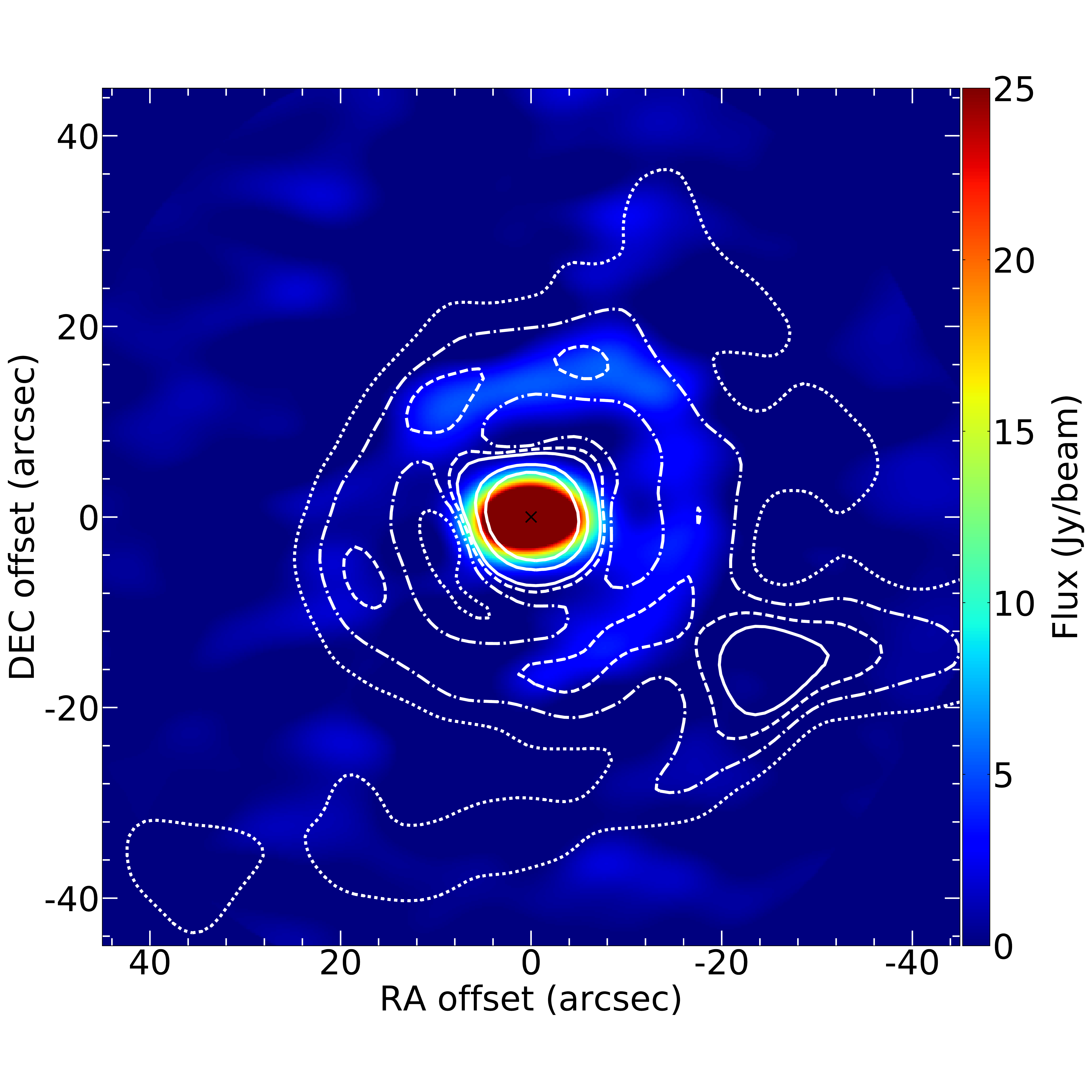} 
   \caption{Integrated intensity (moment 0) map of the CO(2--1) emission observed with the ACA (colourscale) overlaid with contours of the \textit{Herschel/PACS} dust emission. North is up, east is left. The stellar position is marked with a black cross. The contours are plotted in white at 1.1\,$\sigma$ (dotted lines), 1.2\,$\sigma$ (dash-dotted lines), 1.3\,$\sigma$ (dashed lines), 1.4\,$\sigma$, 2\,$\sigma,$ and 3\,$\sigma$ (solid lines). }
   \label{fig:herschelACA}
\end{figure}

We compared the molecular CO(2--1) gas emission, as observed with the lower resolution ACA, with the thermal dust emission at 70\,$\mu$m observed by \textit{Herschel/PACS}. In Fig.\,\ref{fig:herschelACA} we show an integrated intensity (moment 0) map of the ACA observations overlaid with contours of the \textit{Herschel} observations. The gaseous and dusty emissions seem to coincide very well with each other, especially the shell structure is very similar in the two components. The emission blob, picked up with \textit{Herschel} just outside the south-west of the shell, is not particularly visible in the ACA observations. In this region we already expect to observe the ISM interaction front \citep{Jorissen2011}. In general, it is expected that the interstellar radiation field is destroying CO molecules through photo-dissociation at the ISM interaction front, which is consistent with the lack of clear CO detections on these spatial scales.

\section{Discussion}
\label{sec:discussion}

Only a small number of detached shells around AGB stars have been known until now, and while some show a clumpy or irregular fine structure similar to that of TX~Psc \citep[e.g.][]{Kerschbaum2017}, all of them show spherical symmetry when it comes to the general shape of the detached shell. The detached shell around TX~Psc is the first with an elliptical shape. In the subsections below we discuss processes that might significantly influence the shaping of the detached shell, and potentially explain the ellipticity.

\subsection{Stellar motion and ISM interaction}
Although the minor axis of the elliptical detached shell is aligned with the stellar proper motion through space, we can confidently claim that the ellipticity is not caused by ISM interaction. The ISM interaction front can clearly be seen in the \textit{Herschel} dust image \citep{Jorissen2011}, and it lies beyond the detached shell. Furthermore, the shell is also elliptical in the direction opposite to the space motion, and furthermore the present-day wind shows an ellipticity with the same orientation as the shell. Therefore we conclude that the ellipticity of the detached shell around TX~Psc is not caused by external interaction with a different medium, but has to originate from internal stellar properties or processes much closer to the star itself. 

\subsection{Binary companion}
Optical radial velocity measurements of TX~Psc by \citet{Barnbaum1992b,Barnbaum1992a} indicate only minor radial velocity variations, smaller than 5\,km/s, and they even use TX\,Psc as a standard star for the optical radial velocity analysis of their sample. On the other hand, \citet{Jorissen2011} report new observations from 2009-2010 with the HERMES/MERCATOR spectrograph that show slightly larger variations between 8 and 15\,km/s. These variations take place on a timescale much longer than the semi-regular pulsation period of 224\,days, and \citet{Jorissen2011} interpreted these variations as either an influence by an unidentified binary companion, or an example of so-called long-secondary periods \citep{Nicholls2009}. High-resolution studies of TX~Psc achieved from multiple lunar occultations showed that the star cannot be described by a simple stellar disc, but that more likely high-temperature circumstellar dust in a possibly clumpy state, or large cold spots on the stellar photosphere, are responsible for the observed signatures \citep{Richichi1995}. Another interpretation of the lunar occultation data has been published by \citet{Bogdanov1997}, who used a different analysis method to derive the brightness profile, and arrived at the conclusion that TX~Psc has a close binary companion at a separation of 0.052\arcsec and a position angle of 241\deg. GALEX observations in the ultra-violet (UV) wavelength range by \citet{Ortiz2016} also put TX~Psc in the category of possible binary stars since far-UV radiation is detected, which is usually  attributed to a secondary star. Nevertheless, the ratio of the predicted-to-observed near-UV excess does not reach the proposed threshold for a binary indication.

A potential binary companion can significantly shape the stellar wind, but up to now the wind-binary interaction in AGB stars is mainly observed as a spiral structure extending from the star outwards \citep[e.g.][]{Maercker2012,Ramstedt2017,Kim2017}. On the other hand, in the study of planetary nebulae (PNe), the successors of AGB stars, a multitude of shapes and structures of the nebulae has been observed, and a long-standing debate is ongoing about the importance of binaries as sources for most of the observed shapes \citep[e.g. review by][]{De-Marco2009}. Overall, there is no clear consensus on the binary state of TX~Psc in the literature, and we cannot confidently claim binarity from our observations of the CSE alone.

\subsection{Asymmetries in the inner CSE}
On small spatial scales, below the resolution limit of the presented ALMA observations, asymmetries in the inner CSE were detected in multiple studies at multiple wavelengths. \citet{Richichi1995} presented lunar occultation observations in the near infrared and reported that observations in the K-band indicate that the circumstellar emission is more extended along the east-west direction than in the north-south direction. Their average uniform disc diameter is 8.38$\pm$0.05\,mas, but there are substantial variations between the measurements. Adaptive optics observations by \citet{Cruzalebes1998} at the same wavelengths reveal a roughly circular geometry around the star, but an additional clump with a size of 0.25\arcsec and an intensity of 2$\%$ of the total flux in the south-west direction at a distance of $\sim$0.35\arcsec from the star. Interestingly, this clump is located at the same position angle as the potential binary position, reported by \citet{Bogdanov1997}, but significantly farther away from the star. Other observations indicate variable asymmetries on timescales of months \citep{Ragland2006} or even days \citep{Sacuto2011b}. An indication of a bright spot 70-210\,mas south of the star was detected in the near infrared by \citet{Hron2015}.

All of the above listed findings of asymmetry in the close CSE underline the inhomogeneity  of the stellar wind.  We also see these reflected in the clumpiness of the detached shell, but none of the reported results present a valid explanation for the ellipticity of the detached shell.

\subsection{Stellar rotation}

According to theory, AGB stars are believed to be very slow rotators. Despite that, recent observations by \citet{Vlemmings2018} show a directly measured apparent rotation velocity, $v$\,sin\,$i$, of $\sim$1\,km/s for the AGB star \object{R~Dor}. This is still in the regime of low rotation velocities, but two orders of magnitude higher than what is expected from theory. The only exception is \object{V~Hya}, where the fast rotation has been deduced from spectroscopic observations and is most likely explained by angular momentum transfer from a binary companion in a common envelope system \citep{Barnbaum1995rot}. For V~Hya, the authors derive a $v$\,sin\,$i$ of 11\,km/s. Even smaller rotational velocities can lead to a noticeable difference in mass-loss rate and wind velocity between the equatorial plane and the poles, as shown and discussed by \citet{Dorfi1996}. This could lead to an asymmetric and elliptical CSE, such as we observe around TX~Psc.

We used the models and estimates of \citet{Dorfi1996} to calculate the effect of stellar rotation in the case of TX~Psc. Even though modelling of mass loss of AGB stars has improved considerably over the years, this modelling approach still captures the basic processes that determine the magnitudes of the mass-loss rate and gas expansion velocity and thus warrants its application in this context. Small rotational velocities will generate lateral asymmetries in density, temperature, and luminosity. Launching a dust-driven wind at different polar angles can increase these variations, since all dust forming processes depend critically on physical parameters like density and temperature, and the centrifugal force lowers the effective gravity. In order to keep the analysis as simple as possible, we approximated these variations along the polar angle by adopting models for slowly rotating polytropes with a polytropic index of $n=3/2$, which corresponds to a fully convective envelope. Such uniformly rotating polytropes are characterised by a dimensionless parameter 
\begin{equation} \label{e.v_poly}
   \eta= \frac{\omega^2}{2\pi G\rho_{\rm c}}  \:,
\end{equation}
where $\rho_{\rm c}$ denotes the central density and $\omega$ the angular velocity. Adopting a second order expansion in $\eta$, the change of the stellar luminosity and the reduced effective gravity as a result of stellar rotation can be calculated as a function of the polar angle in a coordinate system describing the rotating star. A model with $\eta=0$ corresponds to a non-rotating envelope and $\eta=0.0436$ defines a polytrope rotating at break-up velocity (see e.g. \citet{Tassoul1978} for all details of rotating polytropes). We emphasise that already very small deviations from spherical symmetry with $\eta=0.0004$ in Model {\sf C} (cf.~Table\,\ref{t.am_tx}, last column) are sufficient to generate non-spherical density shells. Adopting these (small) deviations from sphericity, we constructed stationary dust-driven winds within a so-called quasi-spherical approximation, and due to the expanding flow we do not expect any significant lateral interactions. A more detailed description of the models can be found in \citet{Dorfi1996}. 

%

\begin{figure}
   \centering
   \includegraphics[width=0.5\textwidth,trim={0cm 0cm 0cm 0cm}]{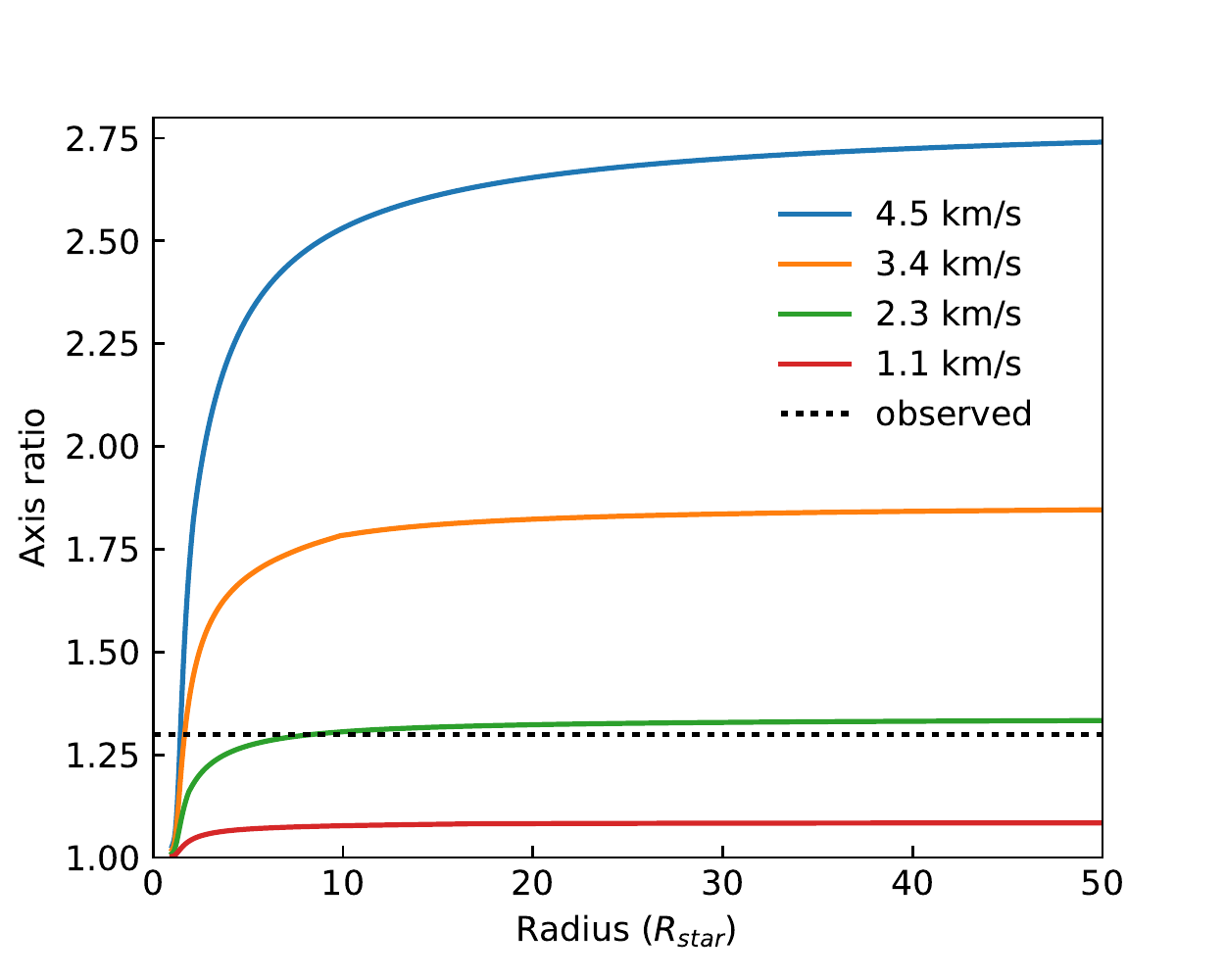}
   \caption{Axis ratio of modelled elliptical density surfaces as a function of radius, shown for different rotational velocities $v\sin i$ (solid lines) and with a stellar radius of $325\,\Rsun$. The observed axis ratio of 1.3 of the elliptical detached shell and present-day wind is plotted as a dashed line.}
    \label{f.den_tx}
\end{figure}

The observed CO-lines are excited by collisions within the outflowing gas, and therefore we can assume that the observed structures are good indications of the gas density structure. Since the wind models are stationary, we have no temporal information on the expanding shell, but we can look for surfaces of constant density as the material is moving outward. The main shaping of non-spherical density shells occurs during the initial acceleration zone within a few stellar radii, as plotted for different rotational velocities in Fig.~\ref{f.den_tx}. Further out, the axial ratio of such density structures becomes constant in the case of a constant outflow velocity. This fact is also confirmed by the observational data, which shows that both elliptical structures seen in the detached shell as well as in the present-day wind exhibit the same axis ratio. For the assumed distance of 275\,pc, the present-day wind has an extent of $\sim 1000\,R_\ast$, which is clearly outside the initial acceleration zone. Using the observed stellar parameters of TX~Psc (see Table~\ref{tab:starparams}) we obtain a mean stellar radius of $R_\ast = 325\,\Rsun$, which was used throughout all our models and calculations. This stellar radius is consistent with the interferometric observations of \citet{Cruzalebes2015}, which yield a stellar radius of 322\,\Rsun and a mass of 2\,\Msun.
We note, that a thermal pulse can synchronise the outflow at different polar angles, but in our stationary models we do not have such a temporal synchronisation that would set a zero point for the flow time. 

The properties of our calculated rotational models are summarised in Table~\ref{t.am_tx}, and the derived axis ratios versus radius of the density surfaces created by the models are plotted in Fig.\,\ref{f.den_tx}. We use the angular velocity $\omega$ as our basic parameter, which corresponds to an observable rotational velocity of $v\sin i = R_\ast \cdot \omega$.
Comparing the different models of Fig.~\ref{f.den_tx}, we see that Model $\sf C$, with a rotational velocity of $\sim 2\,{\rm km/s}$, can explain the observed shell and present-day axis ratio fairly well. We find that already an increase of the best fitting rotational velocity by $50\%$ to $3.4\,{\rm km/s}$ would increase the axis ratio of the shell to a value of $1.84$, which is far beyond the observational values. Lowering the rotation by $50\%$ reduces the aspect ratio to an almost spherical shell, with an axis ratio of $1.08$.

\begin{table}
\caption{Rotational quantities of the TX~Psc models. From the tabulated parameters $v \sin i$ is the rotational velocity, $\omega$ is the angular velocity, $r_e /r_p$ is the ratio of the equatorial axis to the polar axis, $v_{\rm e}/v_{\rm p}$ is the ratio of the equatorial to polar outflow velocity, $v_{{\rm e}, \infty}$ is the final outflow velocity along the equatorial plane, and the last column gives the dimensionless parameter $\eta$, as defined in Eq.~(\ref{e.v_poly}), multiplied by a factor of $10^2$.}
\label{t.am_tx}
\centering
\begin{tabular}{lcccccc}
\toprule
ID
& \multicolumn{1}{c}{\parbox{0.8cm}{\centering $v\sin i$ \\{$({\rm km/s})$}}}         
& \multicolumn{1}{c}{\parbox{1.0cm}{\centering $\omega$ \\{$({\rm s^{-1}})$}}}       
& $r_e/r_p$  
& $ v_e/v_p $ 
& \multicolumn{1}{c}{\parbox{0.8cm}{\centering $v_{{\rm e}, \infty}$ \\{$({\rm km/s})$}}}   
& $10^2\,\eta$ \\
\midrule
{\sf A} & 4.51 & $2.0\cdot 10^{-8}$ & 2.74 & 3.38  & 14.3  & 0.1934  \\
{\sf B} & 3.39 & $1.5\cdot 10^{-8}$ & 1.84 &  2.04 &  12.2 & 0.1088  \\
{\sf C} & 2.26 & $1.0\cdot 10^{-8}$ & 1.33 &  1.42 & 10.1  & 0.0484  \\
{\sf D} & 1.13& $5.0\cdot 10^{-9}$ & 1.08 &  1.10 & 8.5  & 0.0121 \\
\bottomrule
\end{tabular}
\end{table}

Although a rotational velocity of $\sim 2\,{\rm km/s}$ seems small, typical AGB rotation velocities are assumed to be much lower, and a rotation velocity of $\sim 2\,{\rm km/s}$ cannot be achieved without transferring additional angular momentum into the stellar envelope. If we take for example~a mean solar rotational period of $P=24.5\,$days and assume conservation of angular momentum, we calculate a rotational velocity of only $5\,{\rm  m/s}$ when the star expands from $1\,\Rsun$ to $325\,\Rsun$.  

\begin{figure}
   \centering
    \includegraphics[width=0.5\textwidth,trim={1cm 8cm 0cm 8cm}]{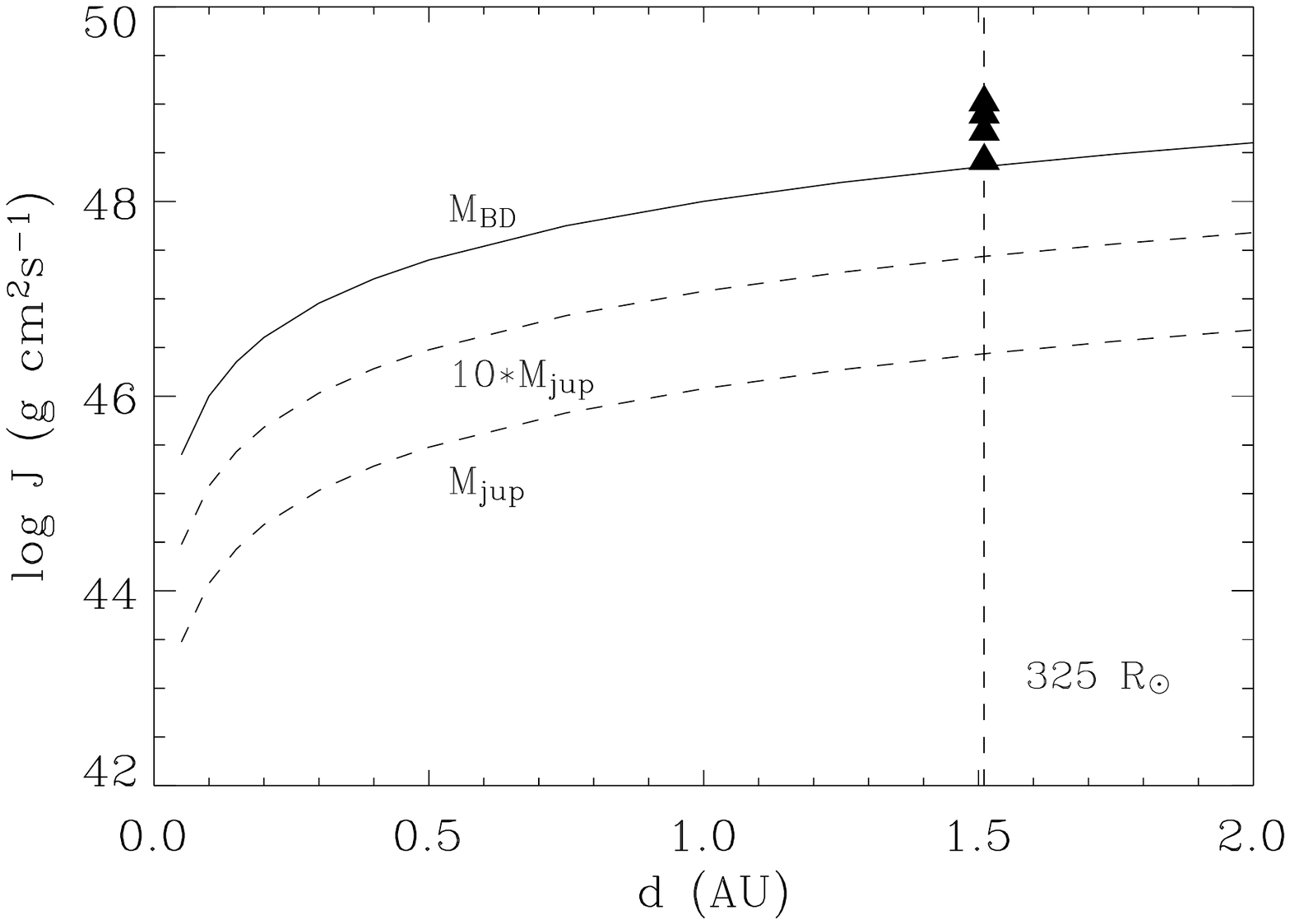}

   \caption{\small Orbital angular momentum for objects of different masses circling around TX~Psc at distances $d$ from the stellar position. The stellar radius of TX~Psc is indicated as a vertical dashed line. The angular momenta of the four rotational models are depicted as black triangles.}
    \label{f.am_tx}
\end{figure}

In Fig.~\ref{f.am_tx} we show the orbital angular momentum of a body of certain mass, circling around a star with a mass of $2\,\Msun$, as a function of distance in AU. We show three different objects: a Jupiter-like object with $M_{\rm jup}$, an object with $10\,M_{\rm jup}$, and a brown dwarf (BD) with $0.08\,\Msun$ ($M_{\rm BD}$, solid line). The vertical dashed line corresponds to the inferred radius of TX~Psc, $\Rtx=325\,\Rsun$. If we assume that the rotational velocity of TX~Psc is generated by an engulfed body, this vertical line will define the outer boundary of such an orbital interaction. The filled triangles are computed for our rotating polytropic models, presented in Table\,\ref{t.am_tx}, and we see that only the body with a BD mass (full line) can meet our requirements for an induced rotational velocity of about $v\sin i \simeq 2\,{\rm km/s}$,  necessary to explain the ellipticity of the observed CO shell.\\ 
The total moment of inertia of TX~Psc, $I_{\rm TX~Psc}$, is difficult to estimate. However, adopting the same model of a slowly rotating polytrope with a polytropic index of $n=3/2$ as used in the above calculations and models, the corresponding moment of inertia $I$ can be calculated analytically (see \citet{Tassoul1978}, p.~246). For a uniformly rotating polytrope with the stellar parameters of TX Psc, we obtain $I_{\rm TX Psc} = 5.2\cdot 10^{56} {\rm g\, cm}^{2}$. We emphasise that due to the small rotation rates, the difference between the equatorial and polar moments of inertia is less than $1\%$. At the level of the second order approximations in $\eta$ (see Eq.\,\ref{e.v_poly}), all quantities describing a fully convective stellar envelope are consistent with the computed polar variations of luminosity, temperature, and density. The plotted values of the angular momentum (filled triangles) are then simply given by
\begin{equation}  \label{e.am_tot}
    J = I_{\rm TX~Psc}\cdot\omega,
\end{equation}
with the values of $\omega$ taken from Table~\ref{t.am_tx}. The process of adding angular momentum of a circulating object to the stellar envelope will be rather complicated, and without detailed numerical simulations it remains unclear to what extent the stellar envelope is being forced into rotation. Hence, we can conclude that our estimates can be used as upper limits to the total angular momentum, since we have assumed a uniform rotation of the whole convective envelope. In a scenario where only part of the envelope has been forced to rotate by the engulfed object, the required mass of the orbiting body can be reduced.

Based on this simple modelling of stationary dust-driven winds of rotating AGB stars, equatorial rotation velocities around $v \sin i \simeq 2\,{\rm km/s}$ are found to be sufficient to explain the observed axis ratio of the elliptical detached shell and present-day wind around TX~Psc. It seems plausible that an engulfed BD could add enough angular momentum to the stellar atmosphere to produce the observed asymmetries in the expanding CO envelope (present-day wind and detached shell).

%
%

\subsection{Stellar evolution}

The low C/O ratio of TX~Psc of an average of 1.07 \citep[reference values in Table\,3 of][]{Klotz2013} suggests that the star only recently turned into a carbon-rich AGB star, changing from oxygen rich to carbon rich during one or more recent thermal pulses. We cannot confidently pinpoint this transition to a single event, such as the thermal pulse that created the observed detached shell, but we can speculate about the likeliness and consequences of this thermal pulse being either the very first or one of the first thermal pulses to have altered the photospheric chemistry of TX~Psc. As described in detail in \citet{Mattsson2007TP}, a detached shell can be formed through the combination of an eruptive mass-loss event and the interaction of a slow and a fast stellar wind. If we assume that the wind-wind interaction acts like a "snow-plow effect" shaping the thin detached shell, one can further presume that a detached shell will be shaped more precisely if the pre-TP, interacting wind is more dense. As thermally pulsing AGB stars will undergo multiple thermal pulses until they reach the end of the TP-AGB phase, subsequent detached shells might run into an increasingly denser and also smoother outer CSE, and therefore later TPs might generate detached shells that show less filaments and less irregular fine structure. One possible upper limit for this effect might be the increase of the inter-pulse mass-loss rate to values so high that the density contrast between detached shell and wind becomes almost indistinguishable. This mass-loss rate increase is expected to happen for stars that become more regular pulsators (i.e. Miras), which could explain the reason for the observation that so-far only irregular or semi-regular variables are found to host detached shells.

In contrast to the above-described "snow-plow effect", the first few TPs might generate less well-structured detached shells because they run into a CSE that has not yet been "swept up" by multiple preceding detached shells. These assumptions could be an explanation for the irregular and patchy fine structure of the detached shell around TX~Psc, which could be produced through one of the very first TPs of this star.
In comparison, the well resolved detached shells of R~Scl \citep{Maercker2012,Maercker2016}, TT~Cyg \citep{Olofsson2000}, and U~Ant \citep{Kerschbaum2017} are significantly smoother with a continuous structure. R~Scl has a C/O ratio of 1.34 \citep{Bergeat2005} and although the C/O ratio is only a rough estimate of the evolutionary status of an AGB star and can be subject to many external changes, R\,Scl can therefore, most likely, be considered to be at a later evolutionary stage than TX~Psc. While the detached shell of R~Scl also shows intensity peaks within the thin shell, it is a clearly continuous and connected shell, unlike what we observe for TX~Psc. U~Ant has a C/O ratio of 1.44 \citep{Bergeat2005} and the detached shell is similarly continuous and thin to that of R~Scl. There is a general weakness of its shell emission in the south-west quadrant, which is anti-correlated with the intensity of the dust emission, seen with \textit{Herschel} \citep{Kerschbaum2010}. The scattered light observations by \citet{Maercker2010} show a distinct separation between different shells seen in gas and dust emission. For TX~Psc, however, the gas and dust distribution seem to be co-spatial (see Sect.\,\ref{subsec:dust}), which is also true for R~Scl \citep{Maercker2014}. This could imply that gas and dust are initially well coupled before effects like photodissociation and interaction with the interstellar radiation field change the molecular gas distribution further along the evolutionary path. A more thorough investigation of the evolutionary state of the compared detached shell sources would for example require an analysis of evolutionary tracks for the observed stellar parameters.

\section{Conclusions}
\label{sec:conclusions}

We have discovered the first elliptical detached shell around an AGB star that, most likely, only recently turned carbon rich through one or several thermal pulses. The molecular gas distribution aligns well with the dust distribution observed in the thermal infrared. The detached shell is clearly separated from the ISM interaction front, and the ellipticity is seen also at smaller spatial scales in the present-day wind. Therefore we conclude that the shaping of the elliptical detached shell has its origin at the stellar level.

We have investigated possible mechanisms behind the ellipticity of the shell and come to the conclusion that a stellar rotation of $\sim 2$\,km/s of TX~Psc could lead to a circumstellar elliptical structure, as a consequence of higher mass-loss rate and expansion velocity in the equatorial plane, which resembles the observed one. This would be the first indirect observation of significant AGB rotation influencing the large-scale geometry of CSEs (while slower AGB rotation has been shown recently by \citet{Vlemmings2018}).  A likely requirement for the conservation and increase of angular momentum, needed to drive such significant stellar rotation, is the existence of a companion object of considerable mass up to the mass of a BD, either in a close orbit or being engulfed by the primary star.

Assuming a rotational ellipsoidal structure of the 3D CSE geometry together with an elliptical velocity distribution, we can constrain the inclination of the rotational axis to roughly 100\deg\  through comparison of observations with geometric models and generated projected-velocity maps. The small-scale structure of the detached shell is very filamentary and patchy compared to other observed detached shells, which raises the question whether this could be an effect of the evolutionary status of TX~Psc. Its very low C/O ratio suggests that it is a young carbon star.

\begin{acknowledgements}
This paper makes use of the following ALMA data: ADS/JAO.ALMA\#2015.1.00059.S. ALMA is a partnership of ESO (representing its member states), NSF (USA) and NINS (Japan), together with NRC (Canada) and NSC and ASIAA (Taiwan) and KASI (Republic of Korea), in cooperation with the Republic of Chile. The Joint ALMA Observatory is operated by ESO, AUI/NRAO, and NAOJ. M.B. acknowledges funding through the Abschlussstipendium fellowship of the University of Vienna. F.K. and M.B. acknowledge funding by the Austrian Science Fund FWF under project number P23586. M.Maercker and H.O. acknowledge support from the Swedish Research Council.

\end{acknowledgements}
%
%

\bibliographystyle{aa}
\bibliography{FULL-literature}


%
%
%

\end{document}